\documentclass[preprint,10pt,nofootinbib]{revtex4-1}
\usepackage[paper=letterpaper,margin=1.5in]{geometry}
\usepackage{amsmath,amssymb,amsfonts}
\usepackage{pstricks}
\usepackage{setspace}
\usepackage[varg]{txfonts}
\usepackage[off]{auto-pst-pdf}
\usepackage[]{graphicx}\graphicspath{{../graph/}%
{../FEL/"FEL at saturation (with Alberto)"/graph/}}
\usepackage{epstopdf}
\usepackage{float}
\usepackage{hyperref}
\hypersetup{
  colorlinks,
  citecolor=green,
  linkcolor=blue}

\newcommand{\p}{\partial}

\renewcommand{\vec}[1]{\textnormal{\boldmath$#1$}}

\begin{document}

\title{Wake excited in plasma by an ultra-relativistic point-like bunch}

\author{G. Stupakov}
\affiliation{SLAC National Accelerator Laboratory,
Menlo Park, CA 94025}
\author{B. Breizman, V. Khudik, and G. Shvets}
\affiliation{Department of Physics and Institute for Fusion Studies, The University of Texas, Austin, Texas 78712, USA}

\begin{abstract}
A plasma flow behind a relativistic electron bunch propagating through a cold plasma is found assuming that the transverse and longitudinal dimensions of the bunch are small and the bunch can be treated as a point charge. In addition, the bunch charge is assumed small. A simplified system of equations for the plasma electrons is derived and it is shown that, through a simple rescaling of variables, the bunch charge can be eliminated from the equations. These equations have a unique solution, with an ion cavity formed behind the driver. The equations are solved numerically and the scaling of the cavity dimensions with the driver charge is obtained. A numerical solution for the case of a positively charged driver is also found.
\end{abstract}

\maketitle

%
\section{Introduction}\label{sec:1}
%

Plasma wakefield acceleration (PWFA) in a highly nonlinear ``blowout'' regime~\cite{PWFA_blowout} offers a promising path  toward high-gradient compact future accelerators for numerous applications. Over the last two decades various aspects of physics involved into the complicated dynamics of plasma excited by a  high-charge driver beam moving with a relativistic velocity has been studied. Unfortunately, due to nonlinear nature of the problem, not much can be done analytically beyond the formulation of the governing system of equations for the plasma flow. Usually, these equation are solved numerically, and computer simulations are widely accepted as a main tool to study PWFA~\cite{Nieter2004448,lotov_2003,mori_2002}. 

While computer simulations are indispensable for testing new ideas and making connections to the experiment, there is also a strong need for analytical techniques that can allow for quick estimates and provide  scaling relations for the governing parameters of the problem. Another application of the analytical approach would be a calculation of the longitudinal and transverse wakefields for the witness bunch accelerated behind the driver beam and the subsequent analysis of the beam instabilities. The importance of such an analysis for the feasibility of the PWFA concept has been recently emphasized in Ref.~\cite{Burov:2016fiw}.

The perceived need to create an approximate analytical description of the PWFA process has lead some of the researchers to developing approximate models that incorporate important features observed in simulations~\cite{Kostyukov_etal,Lu_pwfa}. For example, in Ref.~\cite{Lu_pwfa} an equation for the shape of the plasma bubble is derived based on the assumption that the bubble boundary is formed by a single electron trajectory and the plasma current is localized in a narrow sheath near the boundary. While being a useful complement to numerical simulations, these models usually lack the rigor of the fundamental approach and often have an uncertain range of applicability. In particular, it is not clear whether the model of Ref.~\cite{Lu_pwfa} can be used for the calculation of wakefields inside the bubble, as assumed in Ref.~\cite{Burov:2016fiw}.

An attempt to develop a consistent analytical model for calculations of the plasma flow in the blowout regime is made in this paper. We note that a part of the complexity of the problem is associated with the fact that there are several parameters that control the dynamics of the bubble. Two of these parameters define the size of the driver bunch: $\sigma_r k_p$ and $\sigma_zk_p$, where $\sigma_r$ and $\sigma_z$ are the radial and longitudinal dimensions of the driver and $k_p =\omega_p/c = \sqrt{4\pi n_0r_e}$, with $n_0$ the plasma density and $r_e = e^2/mc^2$ the classical electron radius. In our model, we consider the case of a small-size driver beam, $\sigma_r k_p\ll 1$ and $\sigma_zk_p\ll 1$, and formally take the limit $\sigma_r k_p, \sigma_zk_p\to 0$. This limit eliminates the two parameters from the consideration and simplifies our analysis. While this regime may be away from an optimal setup for achieving the highest accelerating gradient, it can be considered as a first approximation to more realistic situations.  

We also assume an ultra-relativistic driver beam moving through the plasma with the speed of light $c$. This leaves only one external dimensionless parameter---the dimensionless charge of the beam $\nu$---in the problem~\cite{barov_2004},
    \begin{align}\label{eq:1}
    \nu
    =
    Nr_ek_p
    =
    \frac{1}{4\pi}
    \frac{Nk_p^3}{n_0}
    ,
    \end{align}
where $N$ is the number of particles in the bunch. The second equality in this relation shows that within a factor of $\frac{1}{3}$ the parameter $\nu$ is equal to the ratio of the number of particles in the bunch to the number of plasma particles in the sphere of radius $k_p^{-1}$. 

To further simplify the problem we focus this paper on the limit of small charges, $\nu\ll 1$, where, as we will show below, a remarkably simple scaling for the main bubble parameters can be derived.

This paper is organized as follows. In Section~\ref{sec:2}, we recapitulate the main equations that govern the plasma flow and in Section~\ref{sec:3} we formulate initial conditions for these equations. In Section~\ref{sec:4}, we develop a simple ballistic model that demonstrates how a bubble is developed when the plasma self fields are neglected. In Section~\ref{sec:5}, the general system of equations of Section~\ref{sec:2} is simplified in the limit $\nu\ll 1$, rescaled and then solved numerically. Various properties of the bubble obtained from the numerical solution are discussed and related to the ballistic model of Section~\ref{sec:4}. In Section~\ref{sec:6}, we present analytical formulas that remarkably well approximate some of the properties of the numerical solution of Section~\ref{sec:5}. In Section~\ref{sec:7}, we derive the longitudinal electric field $E_z$ on the axis of the bubble. In Section~\ref{sec:8}, a numerical solution for the plasma flow behind a positive driver is presented. The main results of the paper are summarized in Section~\ref{sec:9}.

%
\section{Formulation of the problem}\label{sec:2}
%

In this Section, we formulate the equations that describe plasma dynamics behind a point-like driver moving  with the velocity of light along the $z$ axis through a cold plasma of constant density $n_0$. The charge density of the driver in the cylindrical coordinate system is $\rho_\mathrm{dr}=\mp(Ne/2\pi r)\,\delta(z-ct)\,\delta(r)$, where $r$ is the distance from the axis $z$ and the minus sign refers to an electron driver and the plus---to a positively charged one. 

Following the standard convention, we normalize time to $\omega_p^{-1}$, length to $k_p^{-1}$, velocities to the speed of light $c$, and momenta to $mc$. We also normalize fields to $mc\omega_p/e$, the charge density to $n_0e$, the plasma density to $n_0$, and the current density to $en_0c$. Here $e$ is the elementary positive charge. In these dimensionless units, the charge density of the driver beam is given by the following expression:
    \begin{align}\label{eq:2}
    \rho_\mathrm{dr}
    =
    \mp
    \frac{2\nu}{r}
    \delta(\xi)\delta(r)
    ,
    \end{align}
where $\xi=ct-z$. The uniform plasma at rest occupies the region in front of the beam $\xi<0$. Our goal is to find the plasma flow and the electromagnetic field behind the driver, in region $\xi>0$.

Due to the symmetry of the problem, the plasma flow is axisymmetric. In a steady state, when one can neglect the transients associated with the entering through  the plasma boundary, all the fields depend on $z$ and $t$ in combination $ct-z$, that is through the variable $\xi$. From Maxwell's equations, we find the following equations for the non-zero components of the electric and magnetic fields $E_r$, $E_z$ and $B_\theta$ in the cylindrical coordinate system:
    \begin{subequations}\label{eq:3}
    \begin{align}\label{eq:3a}
    \frac{1}{r}
    \frac{\p}{\p r}
    rB_\theta
    &=
    \frac{\p E_z}{\p \xi}
    -
    j_z
    ,
    \\\label{eq:3b}
    \frac{\p }{\p \xi}
    (B_\theta-E_r)
    &=
    -
    j_r
    ,
    \\\label{eq:3c}
    \frac{\p E_z}{\p r}
    &=
    -
    j_r
    ,
    \end{align}
    \end{subequations}
where $\vec j$ is the plasma current density. The equation for $\nabla\cdot\vec E$ reads:
    \begin{align}\label{eq:4}
    \frac{1}{r}
    \frac{\p}{\p r}
    rE_r
    -
    \frac{\p}{\p \xi}
    E_z
    =
    1-n
    .
    \end{align}
These equations are complemented by the continuity equation for the electron flow
    \begin{align}\label{eq:5}
    \frac{\p}{\p \xi}
    n(1-v_z)
    +
    \frac{1}{r}
    \frac{\p}{\p r}
    rnv_r
    =
    0
    .
    \end{align}
In this equation we used $n\vec v$ for the flow density of the electrons, which is valid if at every point the electron flow is characterized by a unique velocity $\vec v$. In general, however, the flow behind the point-like driver may have several streams at a given point, that is there are several values of $\vec v$ at each point in space. In this case, $n\vec v$ in Eq.~\eqref{eq:5} should be replaced by $\sum_s n_s\vec v_s$ where the summation goes over the different values of $\vec v_s$ with the corresponding densities $n_s$. For notational simplicity, we drop the sum sign in what follows, unless it is specifically indicated.

The equations of motion for the plasma electrons in the dimensionless variables are
    \begin{align}\label{eq:6}
    \frac{dp_r}{dt}
    =
    - E_r
    +
    v_z B_\theta
    ,\qquad
    \frac{dp_z}{dt}
    =
    - E_z
    -
    v_r B_\theta
    ,
    \end{align}
where  $p_r$ and $p_z$ are the radial and longitudinal components of the momentum vector. 

Eqs.~\eqref{eq:3}-\eqref{eq:6} should be supplemented by initial conditions for $n$ and $\vec v$ at $\xi=0$. These conditions are derived in the next section.

%
\section{Plasma crossing of the driver field }\label{sec:3}
%

The driver beam propagating through the plasma with the speed of light carries an electromagnetic field that is localized in an infinitesimally thin ``pancake'' region in the $x-y$ plane $\xi=0$. In what follows, we will call the field in this region the ``driver field''. At a first glance, it might seem that due to the thinness of the driver field region and the fact that it moves with the speed of light, the plasma does not have time to modify it, so the field in this region is the same as in the free space:
    \begin{align}\label{eq:7}
    E_r
    =
    B_\theta
    =
    {\cal A}(r)
    \delta(\xi)
    ,
    \end{align}
where
    \begin{align}\label{eq:8}
    {\cal A}(r)=\mp\frac{2\nu}{r}
    .
    \end{align}
This assumption, however, is incorrect. As was shown in Ref.~\cite{barov_2004}, the radial currents induced when electrons cross the driver field modify it in such a way that
    \begin{align}\label{eq:9}
    {\cal A}(r)
    =
    \mp
    {2\nu}
    K_1(r)
    ,
    \end{align}
where $K_1(r)$ is the modified Bessel function of the second kind. Using the asymptotic expressions for function $K_1(x)$, for small distances, $r\ll 1$ ($r\ll k_p^{-1}$ in dimensional units), we recover from~\eqref{eq:9} the free space expression ${\cal A}(r)\approx\mp{2\nu}/{r}$; in the opposite limit, $r\gg 1$, the plasma shields the driver field to exponentially small values, ${\cal A}(r)\propto \mp e^{-r}/\sqrt{r}$.

Plasma electrons from region $\xi<0$ encounter the driver field, receive radial and longitudinal kicks and change their momentum from the initial zero at $\xi=0^-$ to nonzero values $p_{r0}$ and $p_{z0}$ at $\xi=0^+$. The values $p_{r0}$ and $p_{z0}$ can be easily found if one notices that due to an infinitesimally small thickness of the region of the driver field in the $\xi$ direction, at a given radius $r$, one can locally approximate it by a plane electromagnetic wave. A solution for a point charge motion in a plane electromagnetic field can be found in the literature (see, e.g.,~\cite{landau_lifshitz_ecm,Hora_1989}); from this solution it follows that
    \begin{align}\label{eq:10}
    p_{r0}
    =
    -{\cal A}
    ,\qquad
    p_{z0}
    =
    \frac{1}{2}
    {\cal A}^2
    ,\qquad
    \gamma_0
    =
    1
    +
    \frac{1}{2}
    {\cal A}^2
    ,
    \end{align}
where $\gamma_0 = \sqrt{1+p_{r0}^2+p_{z0}^2}$ is the electron energy corresponding to the momentum $(p_{r0},p_{z0})$. Note the singularities in $p_{r0}$ and $p_{z0}$ when $r\to 0$; these singularities are due to our assumption of the point driver (they do not appear in the case of a finite value of the driver radius). As we will see below, they do not cause problems in finding the plasma flow behind the driver.

Integrating the continuity equation~\eqref{eq:5} through the driver field and noting that both $n$ and $v_r$ have finite values at $\xi=0$ (although they are discontinuous at this point) we find that the product $n(1-v_z)$ does not change from $\xi=0^-$ to $\xi=0^+$. This allows us to find the electron density, $n_*$, at $\xi=0^+$, immediately after the driver field region,
    \begin{align}\label{eq:11}
    n_*
    =
    \frac{1}{1-v_{z0}}
    ,
    \end{align}
where $v_{z0} = p_{z0}/\gamma_0$. In the limit $r\to0$, we have $v_{z0}\to 1$ and the density~\eqref{eq:11} also has a singularity at $r=0$.

After crossing the driver field at $\xi=0$ the plasma flow is governed by a self-consistent field that is determined by Eqs.~\eqref{eq:3}. Eqs.~\eqref{eq:10} and~\eqref{eq:11} provide the initial conditions for Eq.~\eqref{eq:5} and the equations of motion~\eqref{eq:6}. Note that the whole system of equations and the initial conditions have only one dimensionless external parameter, $\nu$. While equations in this and the previous sections are valid for arbitrary $\nu$, in the rest of this paper we will focus on the limiting case of a small driver charge, $\nu\ll 1$. We will see that in this limit a simple rescaling of the variables eliminates $\nu$ from the equations and shows a universal pattern of the plasma flow with a bubble whose shape and characteristic parameters are easily found from a simple computational problem.

%
\section{Ballistic approximation for trajectories}\label{sec:4}
%

Before we analyze the full picture of the plasma flow in the next section, we consider here a simpler problem of the flow at small distances behind the driver beam, $\xi\ll 1$. At these distances we can neglect the effect of the plasma self-fields on the trajectories of the plasma electrons. We call this approximation the ``ballistic'' regime of plasma motion; it assumes that the plasma electrons are moving with constant velocities defined by Eqs.~\eqref{eq:10}. It provides an insight into the formation of the bubble and gives an estimate of its dimensions that will be used in the next section.

In the ballistic approximation, particle trajectories in the $r$-$\xi$ plane are straight lines starting at $\xi=0$ with an offset $r_0$ and going at an angle $\arctan [v_{r0}/(1-v_{z0})]$ to the horizontal axis, 
    \begin{align}\label{eq:12}
    \frac{dr}{d\xi}
    =
    \frac{v_{r0}}{1-v_{z0}}
    .
    \end{align}
As was mentioned in the previous section, we are interested in the regime $\nu\ll 1$. Assuming, in addition, that the region of interest is limited by  $\nu\ll r\lesssim 1$, we find that one can use Eq.~\eqref{eq:8} for $\cal A$, and also ${\cal A}\ll 1$.  It then follows from Eqs.~\eqref{eq:10} that the electrons are moving with non-relativistic velocities $v_{r0}\ll 1$ and $v_{z0}\ll 1$. Hence
    \begin{align}\label{eq:13}
    \frac{dr}{d\xi}
    \approx
    v_{r0}
    \approx
    -{\cal A}
    =
    \frac{2\nu}{r_0}
    ,
    \end{align}
and electron trajectories in the ballistic approximation are given by
    \begin{align}\label{eq:14}
    r
    =
    r_0
    +
    2\nu
    \frac{\xi}{r_0}
    .
    \end{align}
In Eqs.~\eqref{eq:13} and~\eqref{eq:14} we have chosen a minus sign specifying an electron driver. 

\begin{figure}[htb]
\centering
\includegraphics[width=0.6\textwidth, trim=0mm 0mm 0mm 0mm, clip]{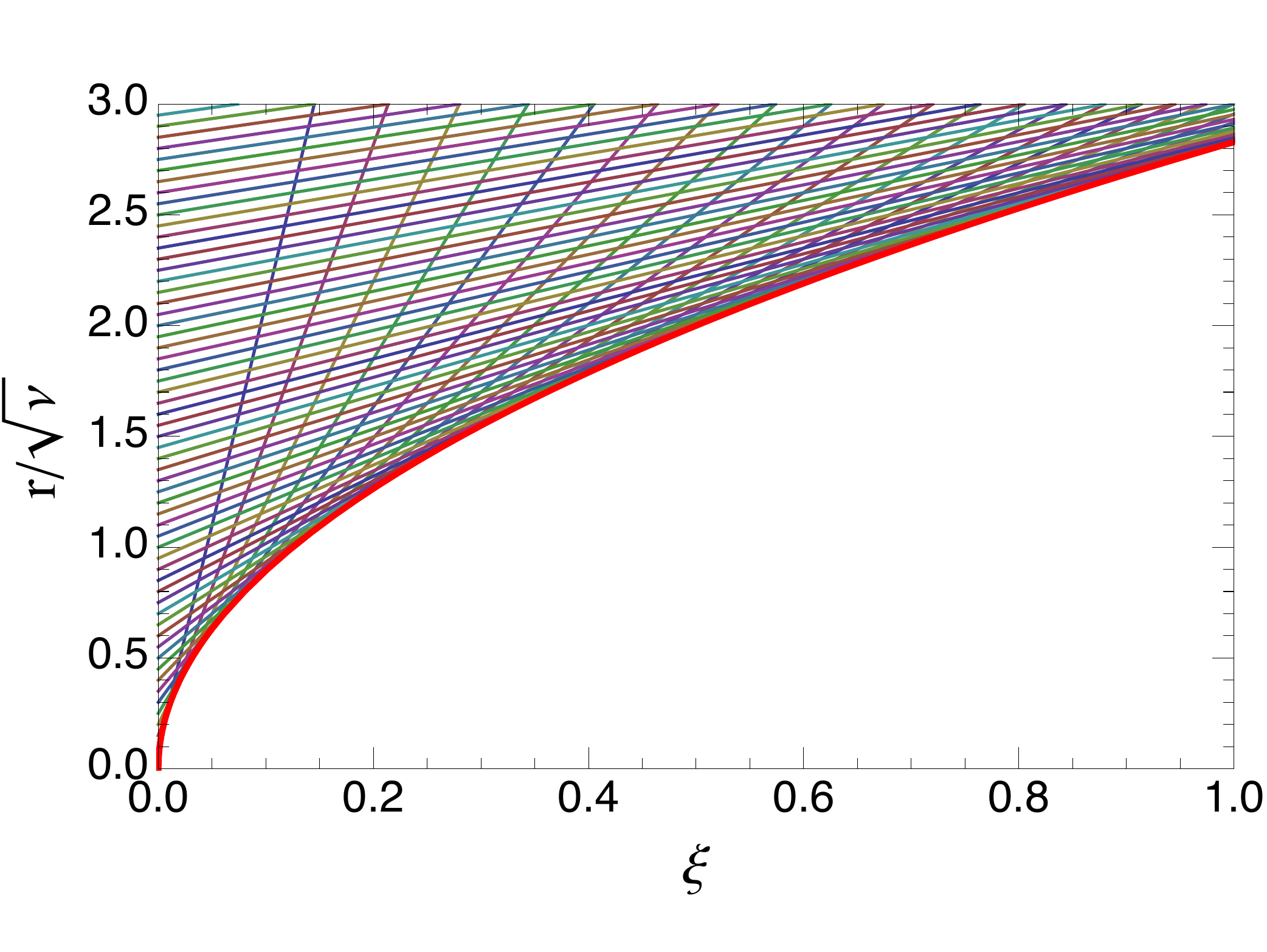}
\caption{Ballistic trajectories in the limit ${\cal A} \ll 1$.  The envelope of the trajectories is shown by the red line. There are no electrons in the region below the envelope.}
\label{fig:1}
\end{figure}

A set ballistic trajectories is plotted in Fig.~\ref{fig:1} where the vertical axis $r$ is normalized by $\sqrt{\nu}$.
With this normalization, the family of the trajectories has a universal shape that does not depend on $\nu$. The plot clearly shows the formation of the bubble boundary  (shown by the red line) below which all the electrons are evacuated and the electron density is zero. An equation for the boundary, $r_b(\xi)$, is found as an envelope for the trajectories with different $r_0$, $dr/dr_0=0$: 
    \begin{align}\label{eq:15}
    r_b(\xi)
    =
    2\sqrt{2\nu\xi}
    .
    \end{align}

As is seen from Fig.~\ref{fig:1}, outside of the bubble there are two electron streams at each point $r,\xi$, corresponding to two different trajectories passing through the point. One of the trajectories touches the bubble boundary after, and the other before, it reaches the point $r,\xi$.

One can also find the density $n(r,\xi)$ of the plasma in region $r>r_b(\xi)$. Derivation of a formula for $n(r,\xi)$ is given in Appendix A.  The density is zero below the bubble boundary, $r< r_b$, and has a singularity $n\propto (r-r_b)^{-1/2}$ as $r\to r_b$ (see Eq.~\eqref{eq:A.5}). 

%
\section{Small-charge regime of the bubble}\label{sec:5}
%

There are several important simplifications that can be carried out in the limit $\nu\ll 1$ as indicated by the analysis of the ballistic approximation in the previous section. First, the region of the nonlinear plasma flow (the bubble and the adjacent region of a large density perturbation), as it turns out, is localized at $r\ll 1$. This allows us to use Eqs.~\eqref{eq:10} with $\cal A$ given by~\eqref{eq:8}; in other words, due to a small transverse size of the flow region, we can neglect the shielding effect in the driver field. Second, a typical value of $\cal A$ in this regime is also small, ${\cal A}\ll 1$, and plasma velocities in the dominant part of the flow are non-relativistic. Using these observations, we replace the initial conditions~\eqref{eq:10} by the following ones:
    \begin{align}\label{eq:16}
    v_{r0}
    =
    \frac{2\nu}{r_0}
    ,\qquad
    v_{z0}\approx 0
    .
    \end{align}

We can obtain a crude estimate of the maximal bubble radius, $r_{bm}$, and the position $\xi_m$ in the $r-\xi$ plane where it is attained  from the following consideration. The radius $r_{bm}$ is where the ballistically streaming trajectories start to bend toward the axis and finally collapse under the influence of the ion focusing field. A plasma electron traveling from $\xi=0$ at $r\sim r_{bm}$ will receive a radial kick  from the ion electric field $E_r$ comparable to its initial radial momentum $p_{r0}$. In dimensional units, $    E_r\sim en_0r_{bm}$, and an electron traveling distance $\xi_m$ with the speed $c-v_z\sim c$ changes its momentum by $e^2n_0r_{bm}\xi_m/{c}$. Equating this to $p_{r0}$ we obtain
    \begin{align}\label{eq:17}
    e^2n_0r_{bm}
    \frac{1}{c}
    \xi_m
    \sim
    mc|{\cal A}|
    ,
    \end{align}
where on the right-hand side we used the initial radial momentum from~\eqref{eq:10}. On the other hand, for an estimate, we can use the relation~\eqref{eq:15} from the ballistic approximation in which $r_b$ is replaced by $r_{bm}$ and $\xi$ by $\xi_m$. From these two equations and Eq.~\eqref{eq:8} for $\cal A$ we find (where we now return to the dimensionless variables)
    \begin{align}\label{eq:18}
    r_{bm}
    \sim
    \sqrt{\nu}
    \end{align}
and $\xi_m    \sim    1$. Because we assume $\nu\ll 1$, we see that the transverse size of the bubble is much smaller than its longitudinal extension, $r_m\ll \xi_m$,---the bubble has a spindle-like shape. Note also that $\cal A$ from~\eqref{eq:8} at $r\sim r_m$ is on the order of $|{\cal A}|\sim\sqrt{\nu}\ll 1$, and hence the plasma electron motion is indeed nonrelativistic, as we have assumed above.

Due to the smallness of the electron velocity we can neglect the magnetic force in Eq.~\eqref{eq:6} and replace it by a simpler one
    \begin{align}\label{eq:19}
    \frac{dv_r}{d\xi}
    =
    - E_r
    ,
    \end{align}
where we used $d/dt=(1-v_z)d/d\xi\approx d/d\xi$. Particle orbits $r(\xi,r_0)$ are determined by $dr/d\xi = v_r$. The electric field in this approximation can be found from Eq.~\eqref{eq:4} in which the term with $E_z$ is neglected as of higher order in parameter $\sqrt{\nu}$,
    \begin{align}\label{eq:20}
    \frac{1}{r}
    \frac{\p}{\p r}
    rE_r
    =
    1-n
    .
    \end{align}
Eqs.~\eqref{eq:19} and~\eqref{eq:20} supplemented by the continuity equation~\eqref{eq:5} (in which $v_z$ in the first term can now be neglected) and the initial conditions~\eqref{eq:16} constitute a full set of equations. An important feature of these equations is that the only external parameter of the problem, $\nu$, can be eliminated by a rescaling of the variables. Indeed, it is easy to see that changing the variables: 
    \begin{align}\label{eq:21}
    r\to \tilde r\sqrt{\nu}
    ,\,\,\,\,
    \xi\to \tilde \xi
    ,\,\,\,\,
    E_r\to \tilde E_r\sqrt{\nu}
    ,\,\,\,\,
    v_r\to \tilde v_r\sqrt{\nu}
    ,\,\,\,\,
    n\to \tilde n,
    \end{align}
does not change the equations but eliminates $\nu$ (replaces it by 1) in Eq.~\eqref{eq:16}.

We solved numerically the rescaled equations by launching a large number $N_\mathrm{tr}$ of trajectories (up to $N_\mathrm{tr}=5000$) at $\xi=0$ with a small radial step $\Delta \tilde r_0$ uniformly distributed from $\tilde r_0=\Delta \tilde r_0$ to $\tilde r_0=N_\mathrm{tr}\Delta \tilde r_0$ (typical $\Delta \tilde r_0\approx 0.003$). The equations of motion for macroparticles were approximated as
    \begin{align}
    \frac{d}{d\xi}
    \tilde v_{r,i}
    &=
    -
    \frac{1}{2}\tilde r_i
    +
    \frac{1}{\tilde r_i}
    \sum_{k=1}^{N_\mathrm{tr}}
    \tilde r_{0,k}
    \Delta \tilde r_0
    \eta(\tilde r_i-\tilde r_k)
    ,\nonumber\\
    \frac{d}{d\xi}
    \tilde r_{i}
    &=
    \tilde v_{r,i}
    ,
    \end{align}
with the initial conditions
    \begin{align}
    \tilde r_i(0)
    =
    \tilde r_{0,i}
    =
    \Delta \tilde r_0 i,
    \qquad
    \tilde v_{r,i}(0)
    =
    \frac{2}{\tilde r_{0,i}}
    ,
    \end{align}
where the macroparticle number $i=1,2\ldots N_\mathrm{tr}$, and $\eta$ is the Heaviside function, $\eta(x)=0$ at $x<0$ and $\eta(x)=1$ at $x>0$. The radial electric field was calculated using Eq.~\eqref{eq:20}, by integrating the electron density over $r$, with the electron density $n$ found from the conservation of the number of macro particles. The equations of motion were solved numerically using an adaptive Runge-Kutta integrator.

\begin{figure}[htb]
\centering
\includegraphics[width=0.6\textwidth, trim=0mm 0mm 0mm 0mm, clip]{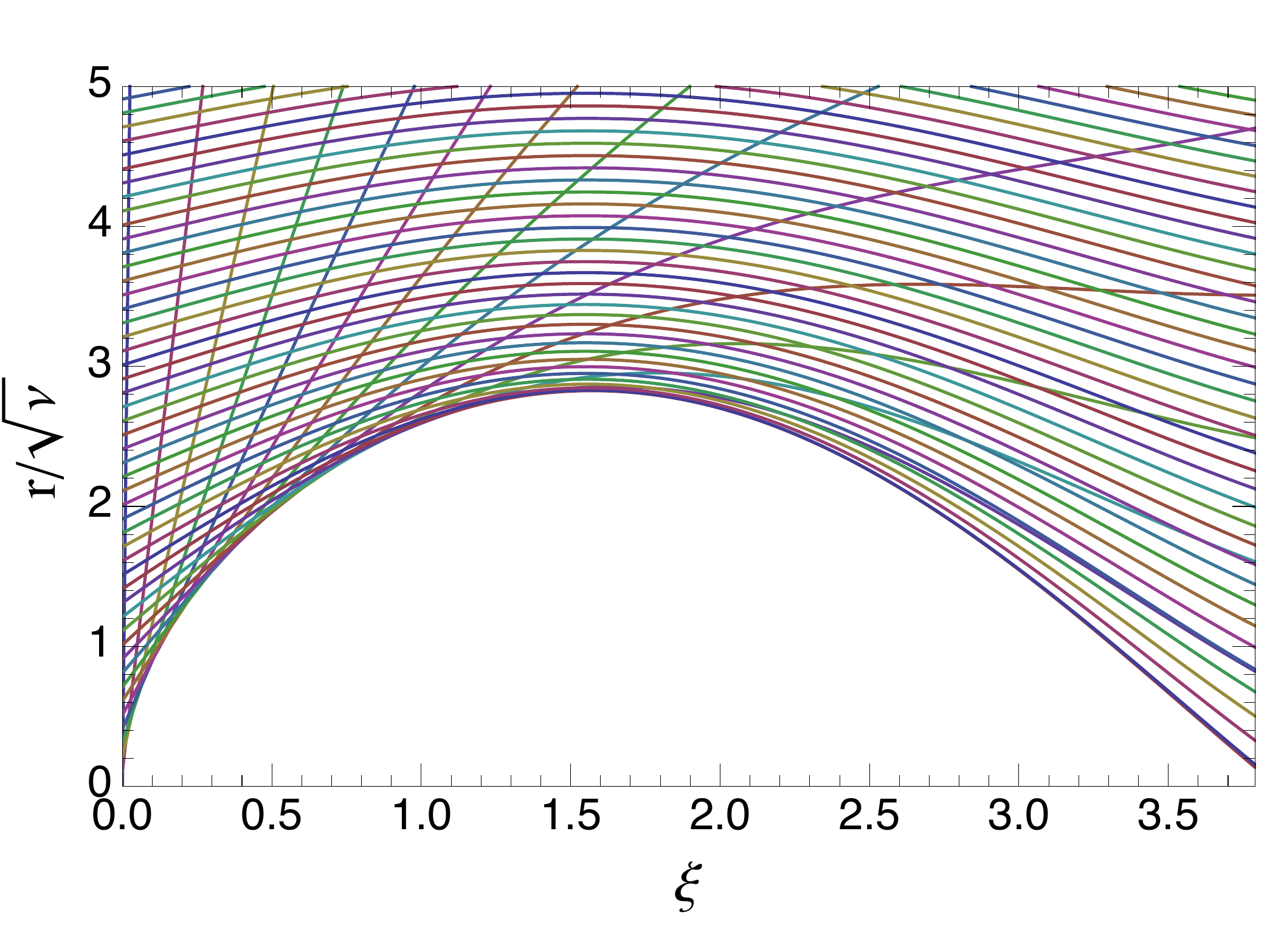}
\caption{Electron trajectories in scaled variables $r/\sqrt{\nu}$ and $\xi$ showing a bubble after the point driver.}
\label{fig:2}
\end{figure}

Fig.~\ref{fig:2} shows the electron trajectories obtained from the numerical solution of the problem. 
The bubble radius (the maximal value or $r$ on the boundary) is reached at $\xi_m=1.57$ and is $r_{bm} = 2.82\sqrt{\nu}$.  The total bubble length (from $\xi=0$ to the point where the bubble boundary collapses on the axis) is $\xi_b = 3.8$ and does not depend on the parameter $\nu$. Note that there are exactly two electron trajectories passing though each point $(r,\xi)$ outside of the bubble which means a two-stream flow at each point. In this regard, the plasma flow is similar to the ballistic model.

\begin{figure}[htb]
\centering
\includegraphics[width=0.6\textwidth, trim=0mm 0mm 0mm 0mm, clip]{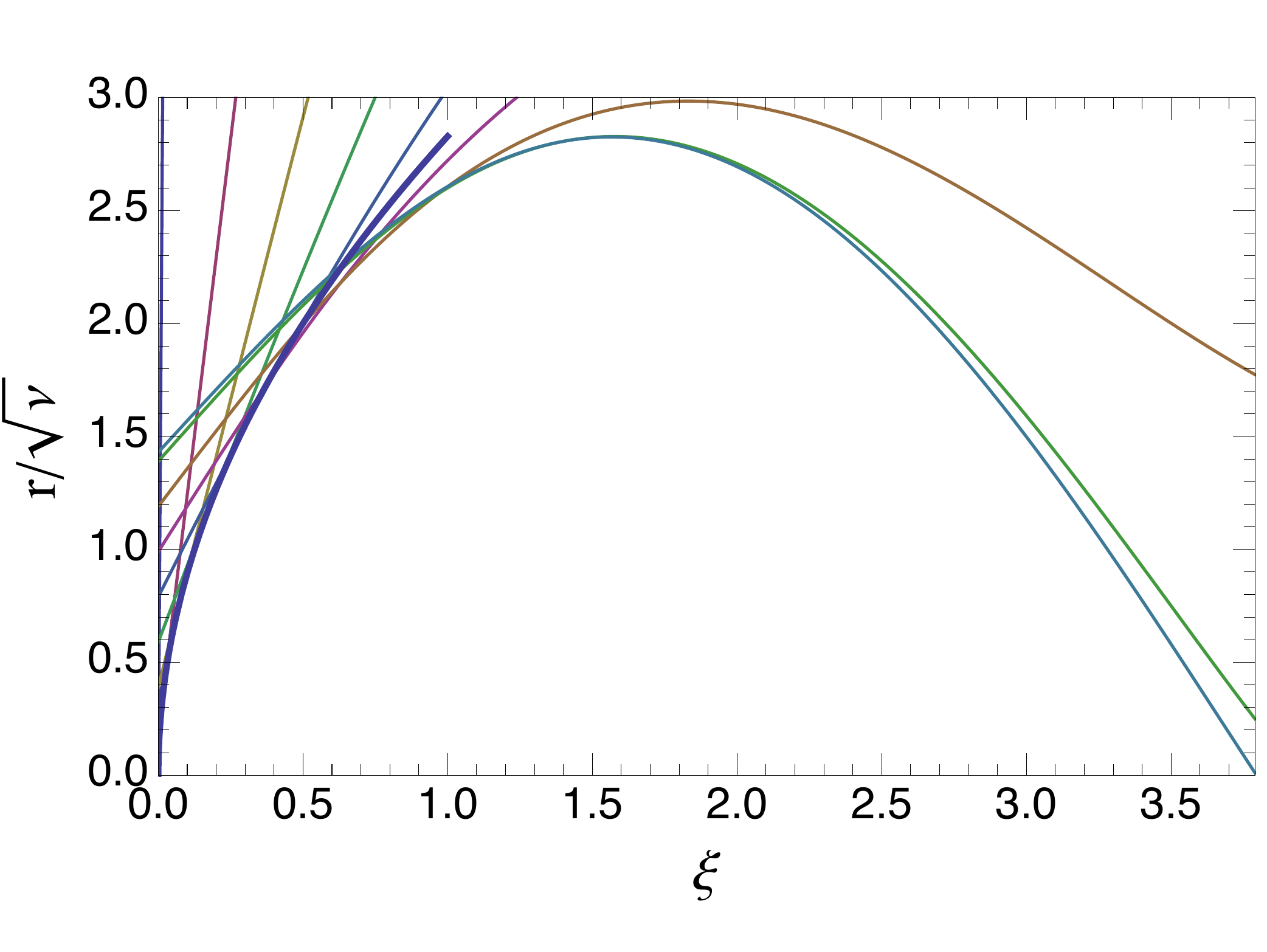}
\caption{A small number of selected trajectories showing a transition from an envelope bubble boundary to a single trajectory. The thick blue line shows the  ballistic approximation, Eq.~\eqref{eq:15},  for the boundary.}
\label{fig:3}
\end{figure}

In the region of not very large values of $\xi$, $\xi\lesssim 0.5$, the trajectories  resemble the ballistic ones from Fig.~\ref{fig:1}. From the analysis of Fig.~\ref{fig:2}, it follows that the bubble boundary before its maximum $r_{bm}$ (that is $\xi<\xi_m$) is comprised of many trajectories for which the boundary is an envelope---the same mechanism of the boundary formation as in the case of the ballistic model from Section~\ref{sec:4}. After the maximum, the boundary is represented by a single trajectory of an electron launched initially at $r_0 =1.44\sqrt{\nu}$. Fig.~\ref{fig:3} shows a small number of trajectories that clearly demonstrate a transition from an envelope structure to a single trajectory boundary (it also shows the boundary calculated in the ballistic approximation).
After the maximum, it is represented by a single trajectory bent toward the axis by the focusing electric field of the ions. 

It is interesting to note that the envelope boundary is different from the assumption made in Ref.~\cite{Lu_pwfa} where a differential equation for the bubble boundary was derived assuming that the boundary coincides with an electron trajectory. Our solution in Fig.~\ref{fig:2} shows that the derivation of~\cite{Lu_pwfa}  is not applicable to our case of a small driver charge and dimensions and hence has a limited range of validity. 

\begin{figure}[htb]
\centering
\includegraphics[width=0.6\textwidth, trim=0mm 0mm 0mm 0mm, clip]{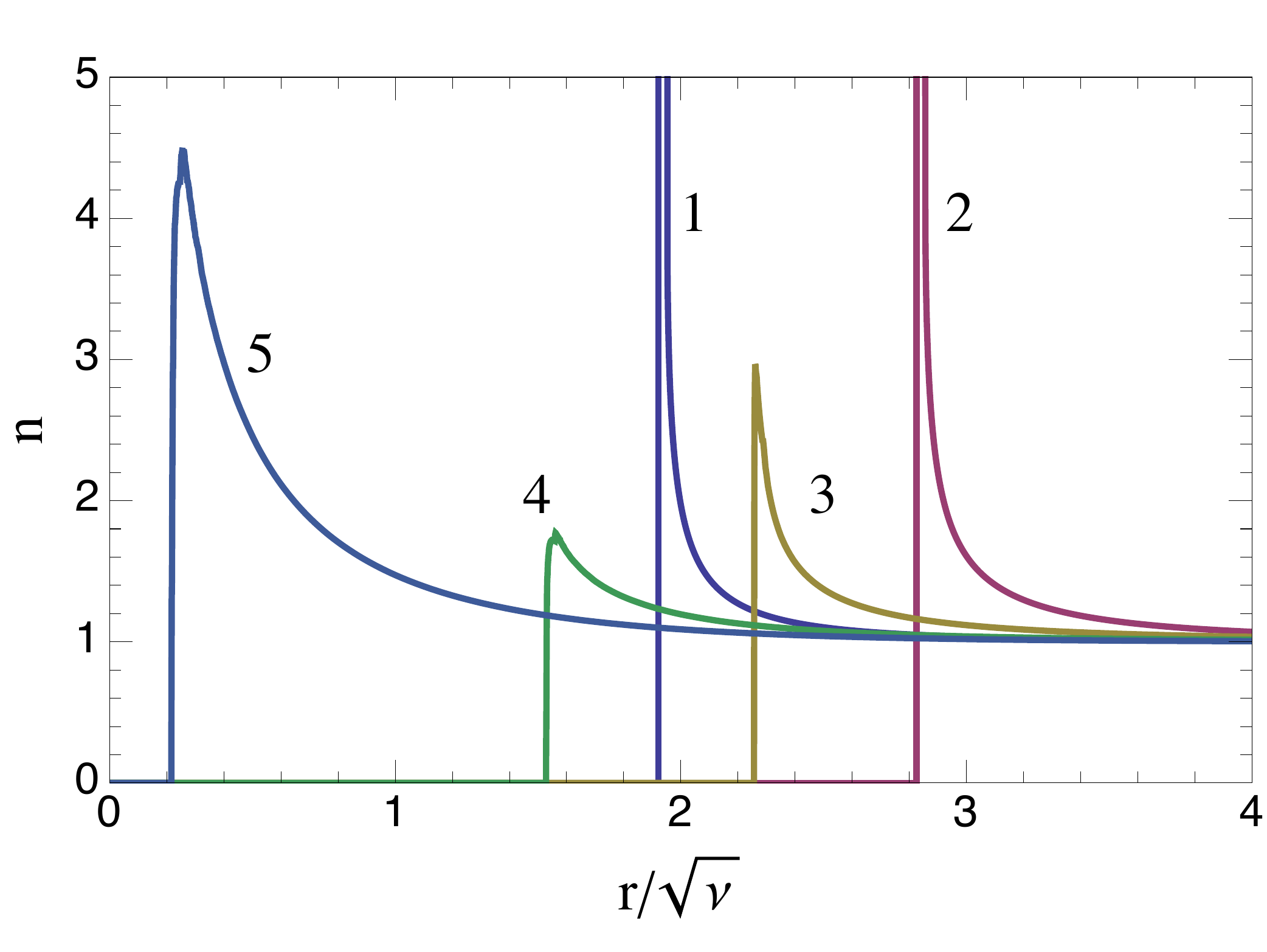}
\caption{Electron density distributions in radial directions for various values of $\xi$: 1) $\xi = 0.5$, 2) $\xi =\xi_m = 1.57$, 3) $\xi = 2.5$, 4) $\xi = 3$, 5) $\xi = 3.7$.}
\label{fig:4}
\end{figure}

Fig.~\ref{fig:4} shows plots of the radial electron density distribution at several values of the longitudinal coordinate $\xi$. 
One can see a zero density inside the bubble with a density jump at the bubble boundary and a gradual fall off to unity as $r\to\infty$. Before the maximum of $r_b$ (curves 1 and 2 in Fig.~\ref{fig:4}) the electron density at the boundary has an infinite value, with the singularity of the same type, $n\propto (r-r_b)^{-1/2}$, as in the ballistic model. After the maximum (curves 3, 4 and 5), the density value on the boundary is finite. Note that the density at the boundary increases from curve 4 to 5 due to the decreasing radius $r_b$; it becomes infinite at $\xi=\xi_b=3.8$ where $r_b=0$.

The fact that the density perturbation in the limit $r\to\infty$ is relatively small, $n-1\ll 1$, indicates that the electron flow at large radial coordinates can be described in linear approximation. We will use this observation in Section~\ref{sec:7} in calculation of the longitudinal electric field inside the bubble.

%
\section{Analytical consideration}\label{sec:6}
%

It seems unlikely that the numerical solution of the previous section can be also obtained analytically, however, we found approximate formulas that remarkably well agree with the numerical results. These formulas are obtained if we replace the ballistic trajectories~\eqref{eq:14} of Section~\ref{sec:4} with electron orbits derived in linear theory. In linear approximation, electrons oscillate with the plasma frequency, and their orbits, in our dimensionless variables, are given by the following equations:
    \begin{align}\label{eq:22}
    r
    =
    r_0
    +
    v_{r0}\sin\xi
    ,\qquad
    v
    =
    v_{r0}\cos\xi
    ,
    \end{align}
where for $v_{r0}$ we use Eq.~\eqref{eq:16}, $v_{r0} = 2\nu/r_0$. Strictly speaking, we expect these equations to be valid at a large distance from the axis, $r\gg 1$, but we will now assume that we can use them within a quarter of the plasma period, $0<\xi<\pi/2$, for all values of $r_0$. In this way, replacing the straight orbits with the oscillating ones, we overcome the deficiency of the ballistic approximation that completely neglects the plasma self field. Note that in the limit $\xi\ll 1$ Eqs.~\eqref{eq:22} reduce to Eqs.~\eqref{eq:14}.

The envelope of trajectories~\eqref{eq:22} is found from the equation $dr/dr_0=0$ which gives
    \begin{align}\label{eq:23}
    r_b(\xi)
    =
    2\sqrt{2\nu
    \sin\xi}
    .
    \end{align}
Using this equation in the interval $0<\xi<\pi/2$ we find for the maximal radius of the bubble and the location of the maximum:
    \begin{align}\label{eq:24}
    r_{bm}
    =
    2\sqrt{2\nu}
    ,\qquad
    \xi_m
    =
    \frac{\pi}{2}
    ,
    \end{align}
in a remarkable agreement with the numerical values for $\xi_m=1.57$ and $r_{bm}=2.82\sqrt{\nu}$ of the previous section.

As was mentioned in the previous section, in region $\xi>\xi_m=\pi/2$, the bubble boundary is a single electron trajectory. This trajectory is easily found from Eqs.~\eqref{eq:19}, \eqref{eq:20} in which we now set $n=0$ (because the electron density is zero inside the bubble). The electric field on this trajectory is $E_r=\frac{1}{2}r$ which gives for the orbit the following equation
    \begin{align}\label{eq:25}
    \frac{d^2 r_{b}}{d\xi^2}
    +
    \frac{1}{2}
    r_{b}
    =0
    .
    \end{align}
Its solution in the region $\xi>\xi_m$ that matches the initial value $r_b(\xi_m)=r_{bm}$ with an addition condition $r_b'(\xi_m)=0$ is
    \begin{align}\label{eq:26}
    r_b(\xi)
    =
    2\sqrt{2\nu}
    \cos
    \left[
    \frac{1}{\sqrt{2}}
    \left(\xi- \frac{\pi}{2}\right)
    \right]
    ,
    \end{align} 
where we have used Eq.~\eqref{eq:24}. As follows from this formula, the bubble boundary collapses on the axis at $\xi = \frac{\pi}{2}(1+\sqrt{2})=3.79$, again in a remarkable agreement with the numerical value for $\xi_{b}$ from the previous section.

\begin{figure}[htb]
\centering
\includegraphics[width=0.6\textwidth, trim=0mm 0mm 0mm 0mm, clip]{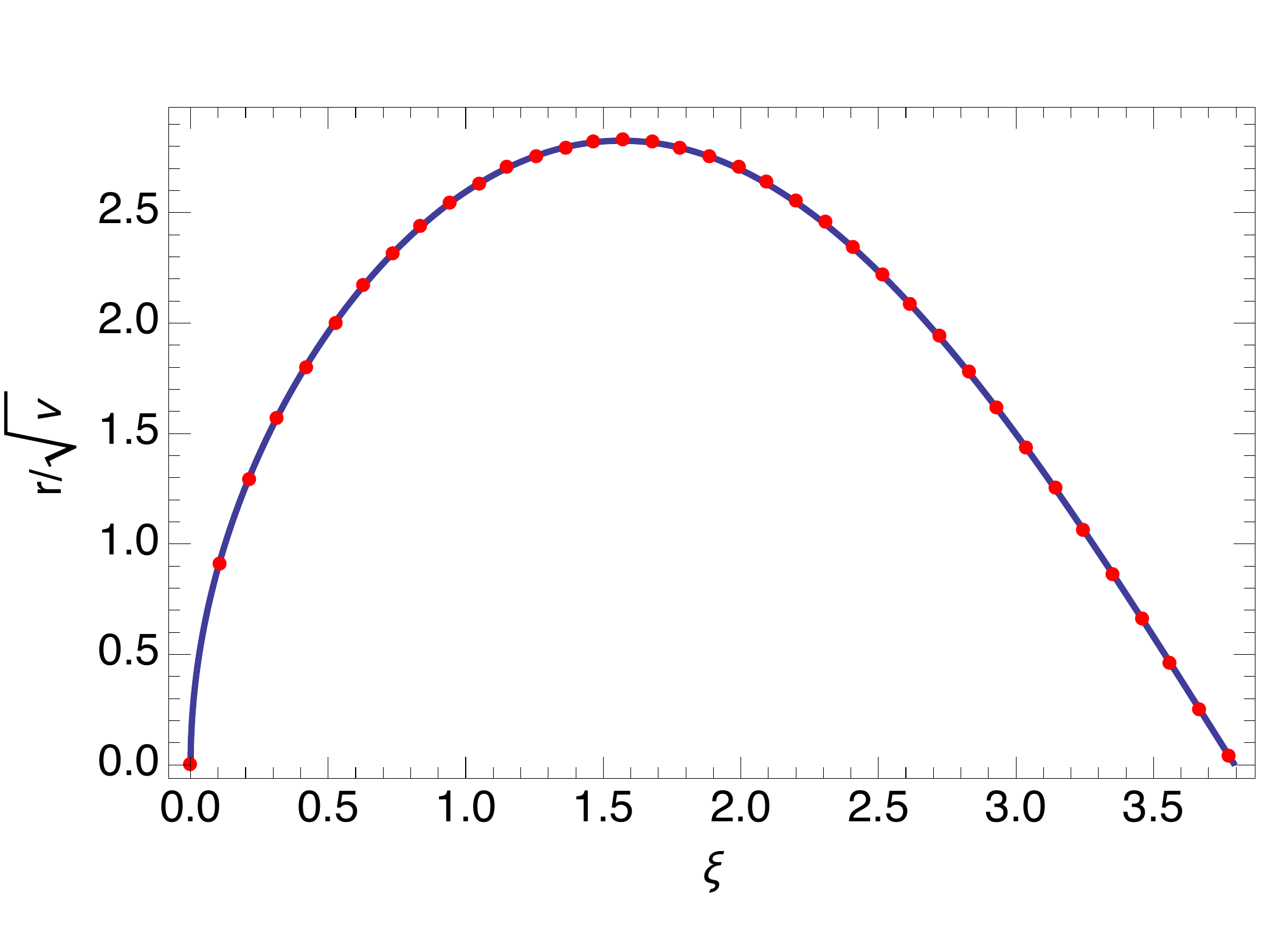}
\caption{Comparison of the analytical expression for the bubble boundary~\eqref{eq:23} and~\eqref{eq:26} (red dots) with the numerical solution (solid line).}
\label{fig:5}
\end{figure}

Comparison of the analytical expressions for the bubble boundary~\eqref{eq:23} and~\eqref{eq:26} with the numerical solution is shown in Fig.~\ref{fig:5}---they agree very well in the whole 
interval $0<\xi<\xi_b$.

%
\section{Longitudinal electric field in the bubble}\label{sec:7}
%

We will now discuss how to calculate the longitudinal electric field $E_z(\xi)$ inside the bubble, on the axis $r=0$. We first integrate Eq.~\eqref{eq:3c},
    \begin{align}\label{eq:27}
    E_z(\xi)
    &=
    \int_{r_{b}}^\infty
    dr'
    \sum_{s=1}^2 n_s(r',\xi)v_{rs}(r',\xi)
    ,
    \end{align}
where we chose the integration constant from the requirement that $E_z\to 0$ when $r\to\infty$. The sum over index $s$ in this equation goes over the two electron streams at a given point $r,\xi$, as discussed in Section~\ref{sec:5}. Unfortunately, an attempt to apply this equation to the numerical solution of the previous section shows that the integral logarithmically  diverges at the upper limit, because $v_r$ at large distances falls off as $1/r$. This happens because we neglected the screening effect of the plasma at $r\gg 1$. To include the screening effect into  calculations, we split the integration interval in~\eqref{eq:27} into two parts choosing some value $r_*$ such that $r_{bm}\ll r_*\ll 1$: the integral from $r$ to $r_*$ is computed using the numerical solution, and the contribution from the region $r>r_*$ is calculated analytically. At large distances, $r> r_*$, we can use a linear theory of plasma oscillations because the initial velocity perturbation $v_{r0}$ is small. In the linear theory, $v_r$ oscillates with the plasma frequency (cf. Eq.~\eqref{eq:22}),
    \begin{align}\label{eq:28}
    v_r
    =
    v_{r0}\cos\xi
    \approx
    2\nu K_1(r)\cos\xi
    ,
    \end{align}
where we have used in the expression for the velocity the value of $\cal A$ given by Eq.~\eqref{eq:9}, thus including the effect of the screening. Since the velocity $v_r$ is small, we can neglect the density perturbation and replace $n$ by unity which gives for the contribution to $E_z$
    \begin{align}\label{eq:29}
    2\nu\cos\xi
    \int_{r_*}^\infty
    dr
    K_1(r)
    =
    2\nu
    K_0(r_*)
    \cos\xi
    \approx
    2\nu
    \left(
    \ln \frac{2}{r_*}
    -
    \gamma_E
    \right)
    \cos\xi
    ,
    \end{align}
where $K_0$ is the modified Bessel function of the second kind, $\gamma_E\approx 0.577$ is Euler's constant and we have used an asymptotic expression for $K_0$ in the limit $r_*\ll 1$. Hence we obtain
    \begin{align}\label{eq:30}
    E_{z}(\xi)
    &=
    \int_{r_{b}}^{r_*}
    dr'
    \sum_{s=1}^2 n_s(r',\xi)v_{rs}(r',\xi)
    +
    2\nu
    \left(
    \ln \frac{2}{r_*}
    -
    \gamma_E
    \right)
    \cos\xi
    .
    \end{align}
We now introduce  $R=r_*/\sqrt{\nu}$; since $r_{bm}=2.82\sqrt{\nu}$, choosing $R\gg 1$ makes $r_*\gg r_{bm}$. We then write
    \begin{align}\label{eq:31}
    E_z = E_{z1}+E_{z2}
    ,
    \end{align} 
where
    \begin{align}\label{eq:32}
    E_{z1}(\xi)
    &=
    \int_r^{R\sqrt{\nu}}
    dr'
    \sum_{s=1}^2 n_s(r',\xi)v_{rs}(r',\xi)
    +
    2\nu
    \left(
    \ln \frac{2}{R}
    -
    \gamma_E
    \right)
    \cos\xi
    ,
    \end{align}
and
    \begin{align}\label{eq:33}
    E_{z2}(\xi)
    &=
    -
    \nu
    \ln \left({\nu}\right)
    \cos\xi
    .
    \end{align}
The contribution $E_{z1}$ is computed  by solving the plasma flow in the region $0<r<R\sqrt{\nu}$ and numerically calculating the integral in~\eqref{eq:32} using this solution. As follows from the derivation, $E_{z1}$ actually does no depend on $R$: the change in the integral due to the variation of $R$ is compensated by the logarithm in the second term in~\eqref{eq:32}. Moreover, looking at the scalings~\eqref{eq:21} we conclude that $E_{z1}$ is linearly proportional to $\nu$. Hence, it can only be computed once, and then used for different values of $\nu$ by a simple rescaling. The second term, $E_{z2}$, is nonlinear in $\nu$, but has an explicit simple analytical form.

\begin{figure}[htb]
\centering
\includegraphics[width=0.6\textwidth, trim=0mm 0mm 0mm 0mm, clip]{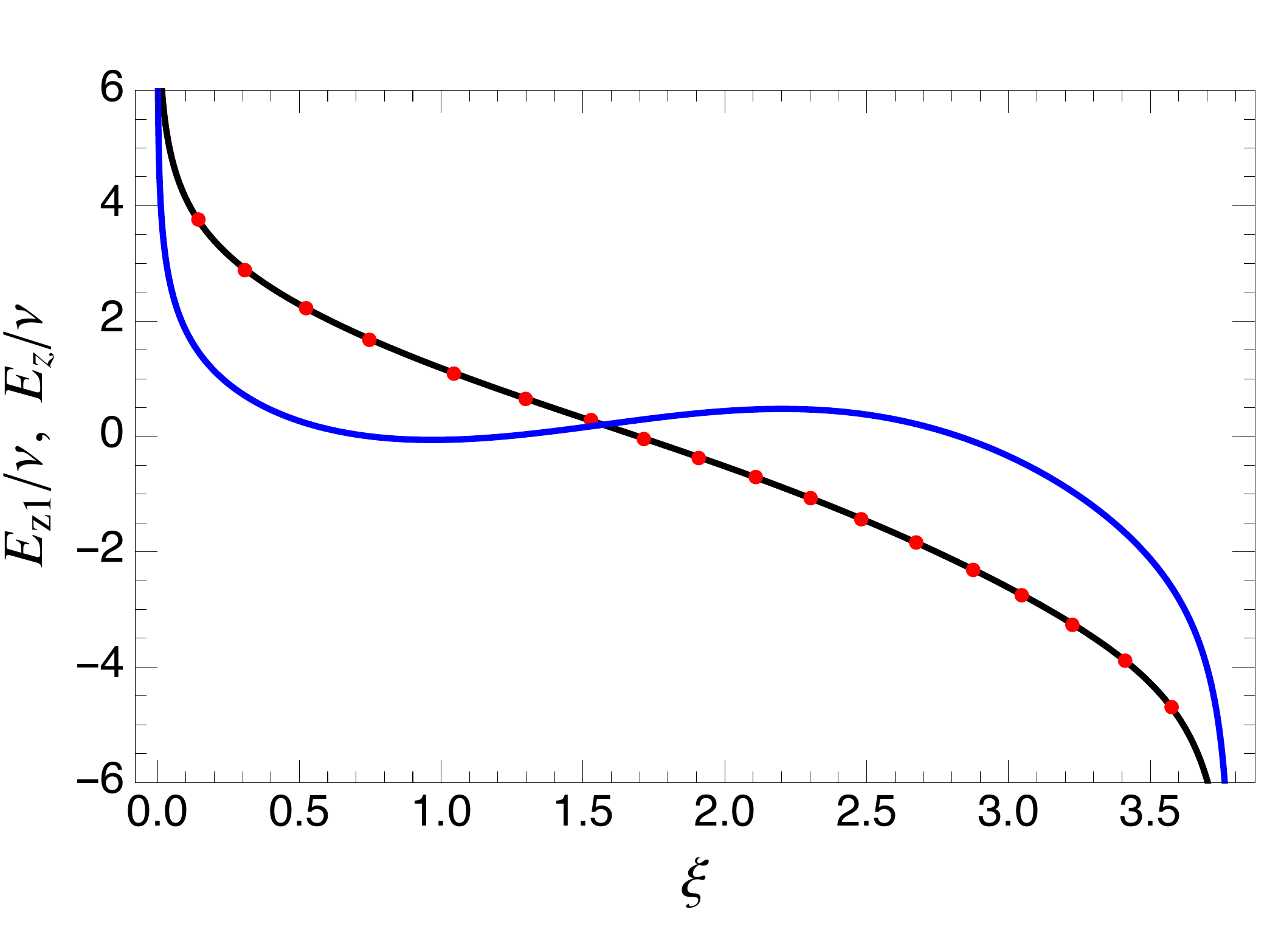}
\caption{Electric field component $E_{z1}$ (blue line) and the total field $E_{z}$ (black line) on the axis of the bubble. The plot of $E_z/\nu$ corresponds to $\nu=0.1$. The fields are  normalized by parameter $\nu$.}
\label{fig:6}
\end{figure}

To illustrate the calculation of $E_z$ we show in Fig.~\ref{fig:6} the plots of $E_{z1}/\nu$ and $E_z/\nu$ calculated with the help of Eqs.~\eqref{eq:31}-\eqref{eq:33}. Note that $E_{z1}/\nu$ does not depend on $\nu$ while $E_z/\nu$ has a part~\eqref{eq:33} that logarithmically depends on $\nu$; the plot of $E_z/\nu$ in Fig.~\ref{fig:6} corresponds to $\nu=0.1$. The black line shows $E_z$ calculated with $R=40$, and the red dots indicate the result of the calculation of $E_z$ with $R=20$.
The agreement between these two calculations corroborate the statement that Eq.~\eqref{eq:32} uniquely defines $E_{z1}$ independent of the value of $R$.

One can see from Fig.~\ref{fig:6} that the field $E_{z}$ changes sign at $\xi=1.69$, close to, but somewhat behind, the location of maximum value of the bubble radius. The sign of  $E_{z}$ is such that electrons inside the bubble would be decelerated at $\xi<1.69$ and accelerated in the region $1.69<\xi<\xi_b=3.8$.

%
\section{Positive driver beam}\label{sec:8}
%

Equations of Section~\ref{sec:5} can be also used to simulate a positive driver beam of small dimensions with $\nu\ll 1$ by changing the sign of $v_{r0}$ in Eq.~\eqref{eq:16}. All the subsequent equations in that section remain valid, including the scaling of the variables~\eqref{eq:21} that eliminates $\nu$ from the equations.

We solved numerically the rescaled equations for the positive driver by the same method as for the electron driver---launching a large number of trajectories uniformly distributed over the initial radial position $r_0$ at $\xi=0$ and then tracing them with account of the electric field of the plasma. An additional complication in numerical algorithm in comparison with the electron driver was caused by the fact that many trajectories cross the axis of the system generating a density singularity on the axis. In the tracking algorithm these trajectories were mirror reflected from the horizontal axis in the $r,\xi$ plane.  

\begin{figure}[htb]
\centering
\includegraphics[width=0.6\textwidth, trim=0mm 0mm 0mm 0mm, clip]{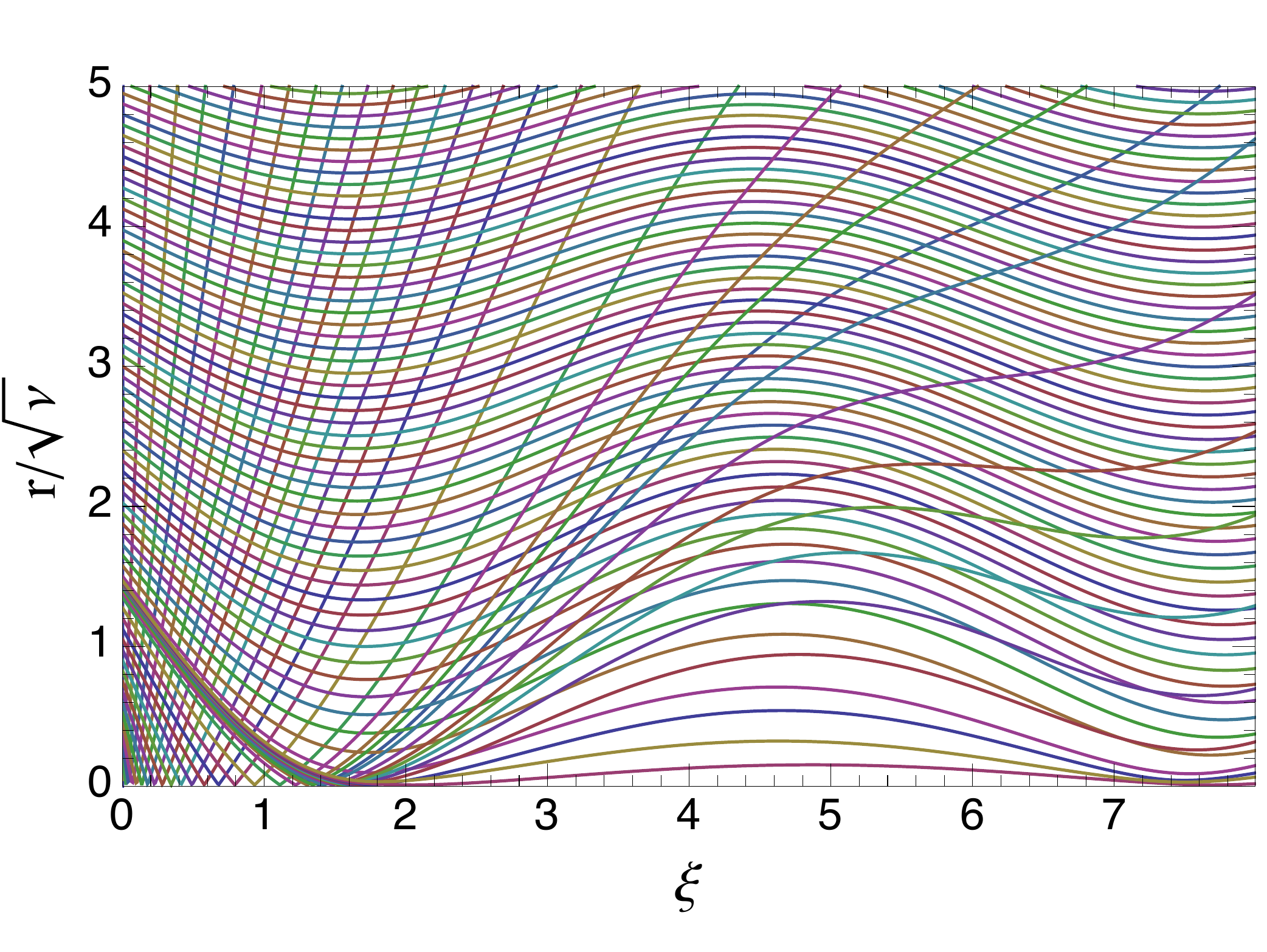}
\caption{Electron orbits in plasma for a positive driver.}
\label{fig:7}
\end{figure}

\begin{figure}[htb]
\centering
\includegraphics[width=0.6\textwidth, trim=0mm 0mm 0mm 0mm, clip]{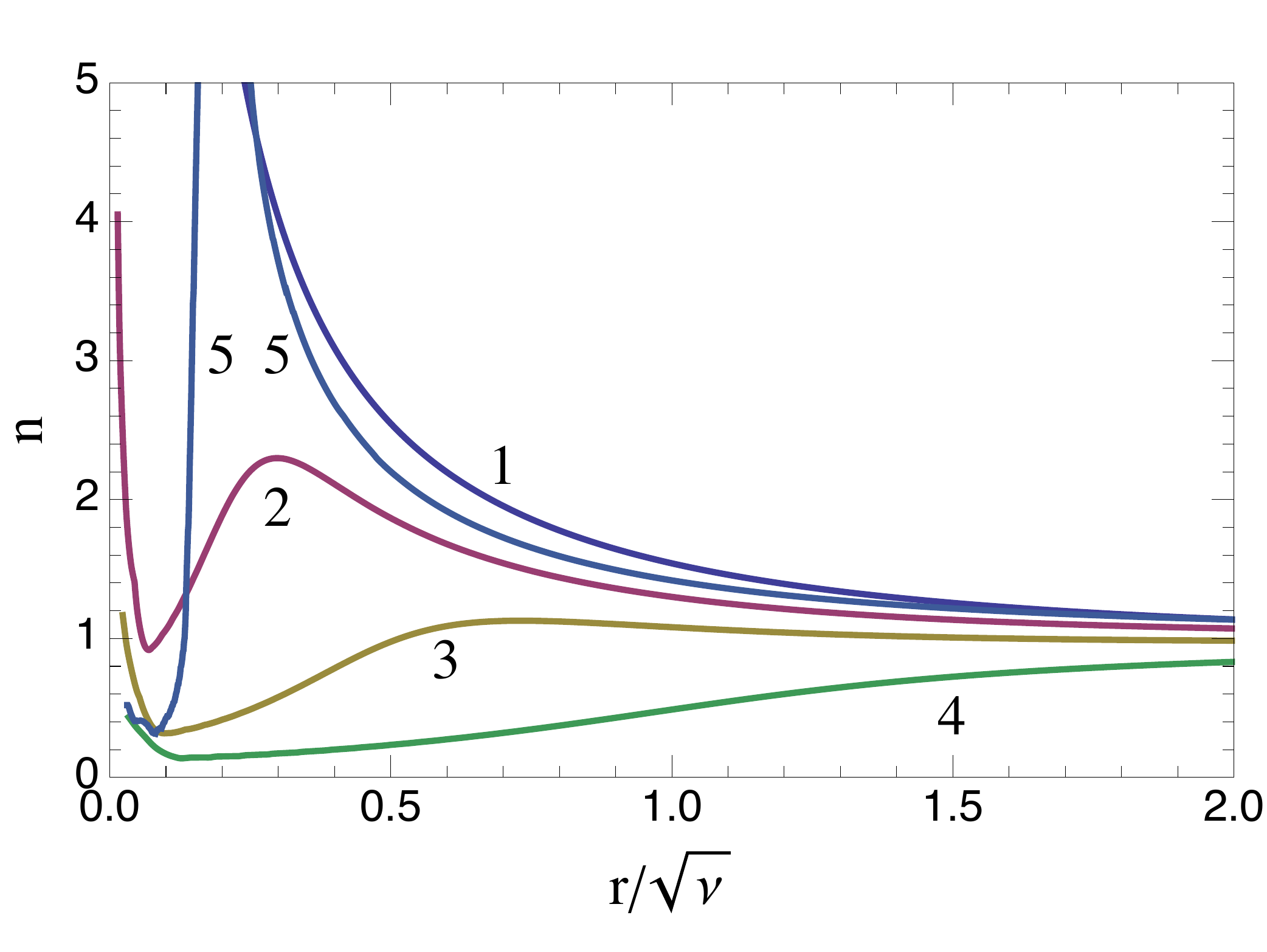}
\caption{Electron density distributions in radial directions for various values of $\xi$: 1) $\xi = 1$, 2) $\xi  = 2.5$, 3) $\xi = 3$, 4) $\xi = 4$, 5) $\xi = 8$. Note that curve 5 has a maximum outside of the plot area and is seen as two light blue lines (each one indicated by number 5).}
\label{fig:8}
\end{figure}

Calculated electron orbits in plasma are shown in Fig.~\ref{fig:7}.
Again, as in the case of the electron driver, there are two electron streams at each point $r,\xi$ with different values of density $n$ and velocity $v_r$. While there is no bubble in this case where electron density is equal to zero, one can see that there is a rarefaction  in plasma trajectories in the region around $\xi\approx 5$, $r\lesssim 1$. This is more clearly seen in the radial density plots shown in Fig.~\ref{fig:8} for several values of $\xi$.
The plasma density in the region $\xi\approx 3-5$ drops considerably below the equilibrium value $n_0=1$ before it reaches a spike on the axis.

%
\section{Summary}\label{sec:9}
%

In this paper we studied the plasma flow behind  a relativistic electron bunch propagating through a cold plasma with the velocity of light. We assumed that the dimensions of the bunch are small so that the bunch can be treated as a point charge. In this model, the governing system of equations for the plasma contains only one parameter, the dimensionless charge of the driver $\nu$, defined by Eq.~\eqref{eq:1}. Assuming additionally that $\nu\ll 1$, we showed that this parameter can be eliminated from the equations, which then have a unique solution. In this solution, for an electron driver, a bubble is formed in the plasma from which electrons are evacuated. We showed that the first part of the boundary of the bubble, before it reaches its maximal radius, is an envelope of many electron trajectories each or which touches the boundary at one point. The electron density has a singularity at this part of the boundary. The second part of the boundary consists of a single electron trajectory, and the electron density is finite at the boundary. Simple analytical expressions were obtained in Section~\ref{sec:6} that describe the shape of the boundary with high precision. We found how the shape of the boundary scales with the driver charge $Q$: its maximal radius $r_{bm}$ scales as $r_{bm}\propto \sqrt{Q}$, and its length does not depend on $Q$.

We also found the longitudinal electric field on the axis of the bubble and showed that it consists of two parts which have different scalings with $Q$: the first one is proportional to $Q$, and the second one scales as $Q\ln Q$.

Finally, a numerical solution of the plasma equations was obtained for the case of a positively charged point driver. This solution does not have a bubble, but shows a region of rarefied electron density at the distance of approximately $5k_p^{-1}$ behind the driver.

%
\section{Acknowledgements}\label{sec:10}
%

G. Stupakov would like to thank Max Zolotorev for useful discussions. 

The work by G. Stupakov was supported by the Department of Energy, contract DE-AC03-76SF00515. The work by B. Breizman was supported by the U.S. Department of
Energy Contract No. DEFG02-04ER54742. The work by V. Khudik and G. Shvets was supported by the U.S. Department of Energy grants DE-SC0007889 and DE-SC0010622, and by an AFOSR grant FA9550-14-1-0045.

\appendix

%
\section{Radial distribution of the plasma density in ballistic model}\label{sec:10}
%

In this section we calculate the plasma density behind the driver in the ballistic approximation of Section~\ref{sec:4}.

\begin{figure}[htb]
\centering
\includegraphics[width=0.35\textwidth, trim=0mm 0mm 0mm 0mm, clip]{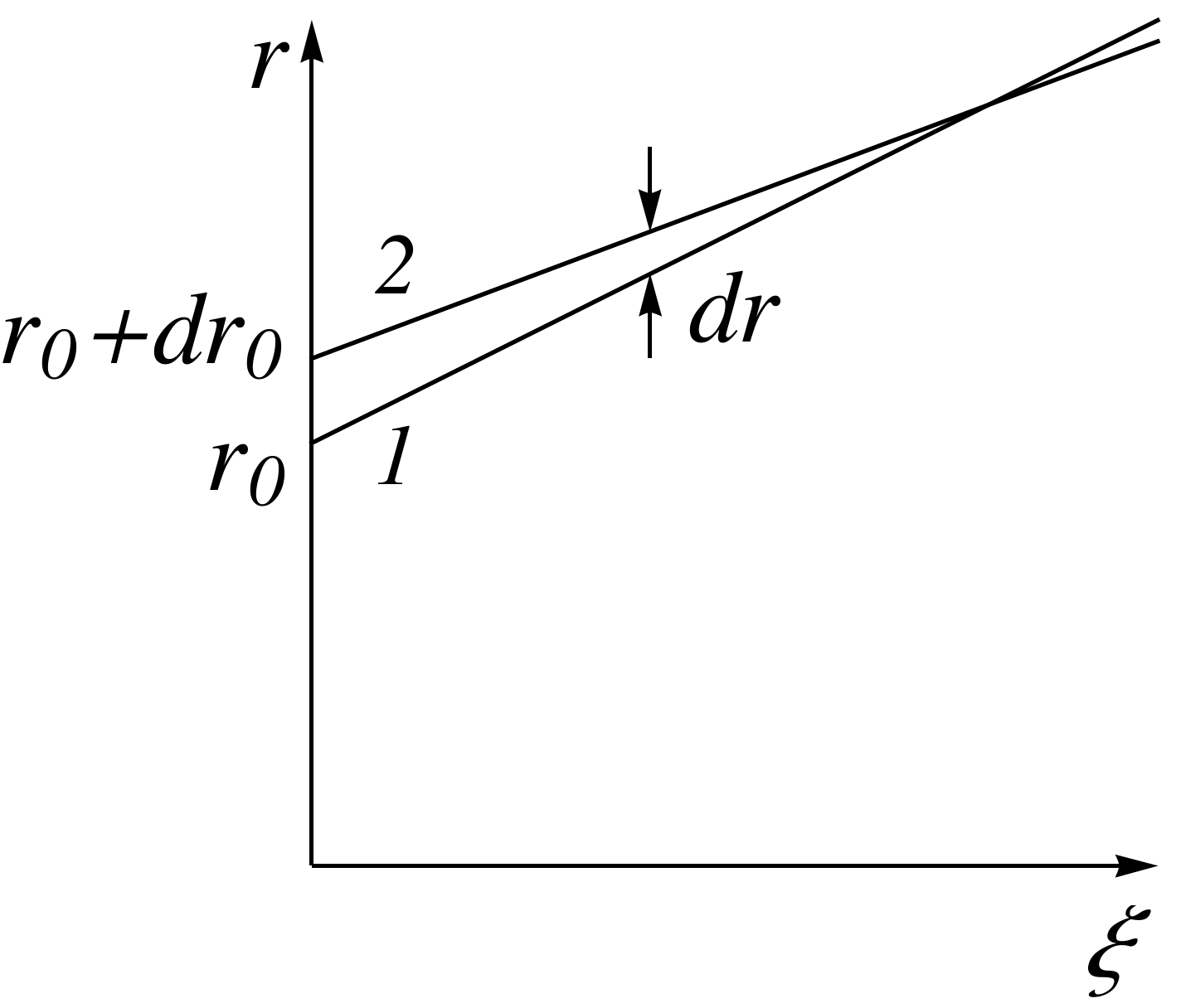}
\caption{Two ballistic electron orbits starting from the initial radial positions $r_0$ and $r_0+dr_0$ after interaction with the driver.}
\label{fig:9}
\end{figure}

Immediately behind the driver, at $\xi=0^+$, the plasma density $n_*$ is given by equation~\eqref{eq:11}. In non-relativistic approximation that we use in this paper, $v_{z0}\ll 1$, the plasma density after crossing the line $\xi=0$ does not change, $n_*=1$. To find the density at point $(\xi,r)$ we use Eq.~\eqref{eq:14} for the electron trajectories. Considering the plasma between two adjacent trajectories 1 an 2 in Fig.~\ref{fig:9}, from  the continuity of the plasma flow we conclude that $n(r,\xi)r\,|dr|=n_* r_0|dr_0|=r_0|dr_0|$, from which it follows that
    \begin{align}\label{eq:A.1}
    n(r,\xi)
    =
    \frac{r_0^3}{r|r_0^2-2\nu \xi|}
    ,
    \end{align}
where we have used~\eqref{eq:14} to calculate $dr/dr_0$. Note that at some value of $\xi$ the density~\eqref{eq:A.1} becomes infinite; this turns out to be on the bubble boundary, as follows from the properties of an envelope of a family of orbits. 

The initial radius $r_0$ in this equation should be expressed through $r$ and $\xi$ from Eq.~\eqref{eq:14}:
    \begin{align}\label{eq:A.2}
    r_0
    =
    \frac{1}{2}
    r
    \pm
    \sqrt{
    \frac{1}{4}
    r^2
    -
    2\nu\xi}
    =
    \frac{1}{2}
    r
    \left(
    1
    \pm
    \sqrt{    1    -    t    }
    \right)
    ,
    \end{align}
where
    \begin{align}\label{eq:A.3}
    t
    =
    \frac{8\nu\xi}{r^2}
    <1
    .
    \end{align}
The two solutions correspond to two trajectories that arrive from different initial radii $r_0$ to a given point $r,\xi$. One of these trajectories arrives before and the other one after it touches the envelope. Correspondingly, at a given $\xi,r$, we need to sum the two densities for both trajectories.

Substituting~\eqref{eq:A.2} into~\eqref{eq:A.1} we obtain
    \begin{align}\label{eq:A.4}
    n_\pm(r,\xi)
    =
    \frac{1}{2}
    \frac{
    \left(
    1
    \pm
    \sqrt{    1    -    t    }
    \right)^3
    }{\big|
    \left(
    1
    \pm
    \sqrt{    1    -    t    }
    \right)^2
    -t\big|}
    .
    \end{align}
For the total density, after some simplifications, we find
    \begin{align}\label{eq:A.5}
    n(r,\xi)
    =
    n_+(r,\xi)+n_-(r,\xi)
    =
    \frac{1}{2}
    \frac{2-t}{\sqrt{1-t}}
    =
    \frac{2r^2-r_b^2}{2r\sqrt{r^2-r_b^2}}
    .
    \end{align}
This density has a square root singularity at the boundary of the bubble.

%


\end{document}